\documentclass[twocolumn]{aastex631}
\usepackage{amsmath,amssymb}
\usepackage{graphicx}
\usepackage{hyperref}
\hypersetup{colorlinks=true,citecolor=blue,linkcolor=blue,urlcolor=blue}
\usepackage{bm}
\usepackage{here}
\usepackage{natbib}
\usepackage{tabularx}
\newcolumntype{Y}{>{\centering\arraybackslash}X} %中央揃え

\begin{document}

\title{Coronal Loops with Different Metallicities and Generalized RTV Scaling Laws}

\author{Haruka Washinoue}
\affiliation{Department of Earth and Space Science, Graduate School of Science, Osaka University, Toyonaka, Osaka 560-0043, Japan}
\affiliation{School of Arts and Sciences, The University of Tokyo, 3-8-1, Komaba, Meguro, Tokyo 153-8902, Japan}

\author{Takeru K. Suzuki}
\affiliation{School of Arts and Sciences, The University of Tokyo, 3-8-1, Komaba, Meguro, Tokyo 153-8902, Japan}
\affiliation{Department of Astronomy, The University of Tokyo, 7-3-1, Hongo, Bunkyo, Tokyo, 113-0033, Japan}

\begin{abstract}
%Stellar coronal loops are a dominant source of the X-ray and ultraviolet radiations that have an impact on planetary environments.
Stellar metallicity is a critical factor to characterize the stellar coronae because it directly affects the radiative energy loss from the atmosphere. 
By extending theoretical relations for solar coronal loops introduced by \cite{Rosner1978}, we analytically derive scaling relations for stellar coronal loops with various metallicities.
In order to validate the derived relations, we also perform magnetohydrodyamic simulations for the heating of coronal loops with different metallicities by changing radiative loss functions according to the adopted elemental abundances.
The simulation results nicely explain the generalized analytical scaling relations and show a strong dependence of the thermodynamical and radiative properties of the loops on metallicity. 
Higher density and temperature are obtained in lower-metallicity coronae because of the inefficient radiative cooling, provided that the surface condition is unchanged.  
Thus, it is estimated that the X-ray radiation from metal-poor coronae is higher because of their denser coronal gas.
The generalized scaling laws can also be used as a tool to study the condition of high-energy radiation around magnetically active stars and their impact on planetary environments.
\end{abstract}

%\keywords{magnetohydrodynamics (MHD) --- Stars : Coronae --- Stars}

\section{Introduction}
\label{sec1}

The late-type stars with a surface convective envelope possess a coronal atmosphere with typically more than one-million Kelvin.
High-energy radiations from stellar coronae are considered to significantly affect the formation and evolution of planetary systems.
In particular, the X-ray and ultraviolet (XUV) radiations from stellar coronae are believed to be one of the main players in dispersal of protoplanetary disks \citep{Gorti2009, Owen2012, Nakatani2018, Komaki2021} and the evaporation of planetary atmospheres  \citep{Lammer2003, Sanz-Forcada2011,Bolmont2017,Mitani2020,Rogers2023}.
The XUV radiation from low-mass Population II/III stars could also affect the structure formation in the epoch of cosmic reionization \citep{Washinoue2021}.
Therefore, it is essential to understand the coronal properties under various stellar environment for addressing these unresolved issues.

Since the identification of the high-temperature solar corona around 1940 \citep{Grotrian1939,Edlen1943}, studies on coronal heating have progressed in the Sun, the nearest star. 
Recently, detailed fine-scale structures of the solar corona can be seen with high-resolution observations such as HINODE \citep[e.g.,][]{Kosugi2007,Reale2007,Katsukawa2007,Kusano2012} and the Solar Dynamics Observatory
(SDO) \citep[e.g.,][]{Pesnell2012,Takasao2012, Cheung2015}.
These have revealed that the XUV emission from the solar corona mostly comes from closed magnetic loops with various lengths \citep{Reale2014}.
As a basic theoretical model, \cite{Rosner1978} constructed the scaling relations between loop quantities in an isolated solar coronal loop.
When the energy equilibrium among heating, radiative cooling and thermal conduction is satisfied in a loop, they found that the maximum temperature $T_{\rm max}$ and heating rate $E_{\rm H}$ are related to coronal pressure $P$ and loop length $l$:
\begin{align}
T_{\rm max} = 1400 (Pl)^{1/3},  
\label{eq1}
\end{align}
\begin{align}
E_{\rm H}=9.4\times 10^4 P^{7/6} l^{-5/6}. 
\label{eq2}
\end{align}
Equations (\ref{eq1}) and (\ref{eq2}) have been widely used to understand the  relation between the thermodynamical properties and the size of solar coronal loops and utilized as heating diagnostics in both observational and numerical studies.

Other than the solar corona, a wide range of stellar coronal activities has been observed through detection of the X-ray emission to date, where the activity level depends on stellar parameters such as age and mass \citep{Pizzolato2003,Vidotto2014,Johnstone2021}.
In particular, with recent progress in planetary science, it has been emphasized the importance of studies on various planetary environments which are significantly affected by XUV radiations from a central star.
As a result, there is a growing need to understand the magnetic activities and radiative properties in a wide range of stars.
Therefore, studies that connect between solar and stellar physics have become increasingly crucial.
Recently, combining a Sun-as-a-star approach and observations of Sun-like stars, \cite{Toriumi2022} \citep[see also][]{Toriumi2022b} proposed a unified picture of stellar coronae.
In addition, the relation between EUV and X-ray luminosities of stellar coronae is discovered with a simple formula that is empirically derived from numerical simulations by \cite{Shoda2021}.

Stellar coronal heating in various late-type stars has also been studied with theoretical approach \citep{Cranmer2011, Sakaue2021, Sakaue2021a}.
Recently, it is found that stellar metallicity is an important factor that determines the coronal properties.
This is pointed out by \cite{Suzuki2018} and  \cite{Washinoue2019} who performed the magnetohydrodynamic (MHD) simulations of coronal heating in different-metallicity stars.
They demonstrated that the cooling efficiency in the coronal atmosphere sensitively depends on metallicity which also affects the XUV luminosities and mass loss rate.

Although these studies extended our understanding for stellar coronal heating, their simulations adopt the limited treatment for heating mechanism.
They consider only the shock heating via nonlinear mode conversion from transverse Alfv\'{e}n waves to longitudinal waves, while multiple heating processes are at work in realistic stellar atmospheres. 
In \cite{Washinoue2019}, because of the intermittency of shock heating, the simulated coronal loops are thermally unstable and the loop structures dynamically change with time. 
However, \cite{Matsumoto2014}, who studies detailed heating processes by numerical simulations for the solar corona, reported that the shock heating has a relatively weak contribution to the total heating compared to the turbulent heating in the upper atmosphere. 
Therefore, it is needed to re-examine the metallicity dependence in thermally-stable loops incorporating the heating from turbulent dissipation in addition to the shock formation.

This study aims to understand the effect of stellar metallicity on coronal heating and characterize magnetic loops with different metallicities.
We present a new theoretical model for stellar coronal loops in forms of extended RTV scaling laws and numerical simulations improved from previous studies.
In addition to the shock heating as the previous studies consider, we phenomenologically treat turbulent heating.

In Section \ref{sec2}, we construct the generalized scaling laws for a hydrostatic coronal loop which take into account a difference in metallicity.
In Section \ref{sec3}, the setting of numerical simulations is presented.
In Section \ref{sec4}, we show the results of our simulations to test the generalized scaling laws presented in Section \ref{sec2}.
We discuss our results in Section \ref{sec5} and summarize the paper in Section \ref{sec6}.

\section{Generalized scaling laws}
\label{sec2}

\begin{figure}[!t]
\begin{centering}
\includegraphics[width=8.5cm]{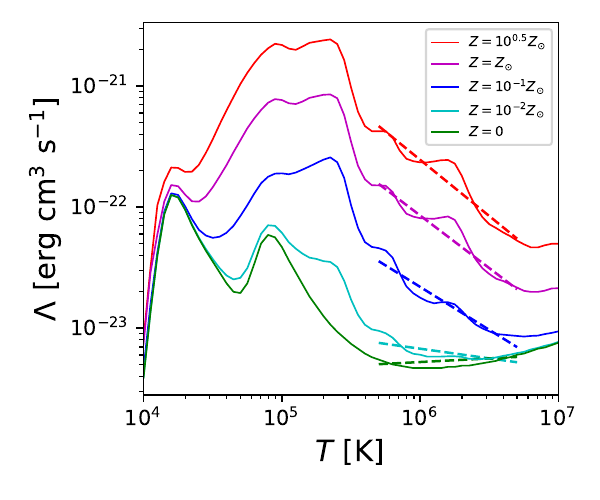}
\end{centering}
  \caption{Optically thin cooling functions for different metallicities from \cite{Sutherland1993}. The dashed lines are the single power-law fitting over $ 10^{5.7}$ K $\leq T \leq 10^{6.7}$ K.}
  \label{fig_fitz}
\end{figure} 

The RTV scaling laws (Equations (\ref{eq1}) and (\ref{eq2})) are derived using the radiative loss function for plasma gas with the solar abundances.
Therefore, they can be only applied to coronal loops in the Sun or stars with the solar chemical composition. 
%To characterize various stellar coronae, it is required to extend the relations to stellar coronal loops with different metallicities.
In this section, we derive generalized scaling laws for coronal loops with different metallicities by a semi-analytical method based on \cite{Rosner1978} and \cite{Hood1979}.
We start from an energy-equilibrium equation among heating, radiative cooling and thermal conduction in a hydrostatic loop:
\begin{align}
E_{\rm H} + E_{\rm R} - \nabla \cdot F_{\rm c} = 0,
\label{eq3}
\end{align}
where $E_{\rm R}$ is the radiative loss rate and $F_{\rm c}$ is the thermal conductive flux.
$E_{\rm R}$ and $F_{\rm c}$ are written as
\begin{align}
E_{\rm R} = -\frac{P^2}{4k_B^2 T^2} \Lambda (T),
\label{eq4}
\end{align}
\begin{align}
F_{\rm c}(s) = -\kappa T^{5/2} \frac{dT}{ds},
\label{eq5}
\end{align}
where $k_B$ and $\Lambda(T)$ are the Boltzman constant and radiative loss function; when deriving Equation (\ref{eq4}), we assumed fully ionized hydrogen plasma, following \cite{Rosner1978}. 
$\kappa$ is the Spitzer conductivity.
We here approximate the cooling functions with the single power-law in  the following form;
\begin{align}
\Lambda(T) = \chi T^{\alpha}.
\label{eq6}
\end{align}
Integrating Equation (\ref{eq3}) from $T_0$ to $T$ using Equations (\ref{eq4})-(\ref{eq6}), we obtain
\begin{align}
F_{\rm c}^2(T) - F_{\rm c}^2(T_0) = \frac{\kappa P^2}{2k_B^2} \int_{T_0}^T dT_1 \chi {T_1}^{\alpha+1/2} - 2\kappa \int_{T_0}^T dT_1 {T_1}^{5/2} E_{\rm H}
\label{eq7}
\end{align}
$T_0 = 2 \times 10^4$ K is set at the bottom of the transition region since we only consider the coronal part in the loop. 
Then, $F_{\rm c}^2(T_0)$ is negligibly small relative to the other terms in Equation (\ref{eq7}).
We also assumed a constant pressure $P$ over the coronal part because the loop length is generally smaller than one pressure scale height of the coronal-temperature gas except for very long loops.
Integrating Equation (\ref{eq7}) from the bottom boundary to the loop top after replacing $F_{\rm c}(T)$ by Equation (\ref{eq5}) yields
\begin{align}
\begin{split}
l = \kappa \int_{T_0}^{T_{\rm max}} &dT_1 {T_1}^{5/2} \left[ \frac{\kappa P^2}{2k_B^2} \int_{T_0}^{T_1} dT_2 \chi {T_2}^{\alpha+1/2} \right.\\
&\left.-2 \kappa \int_{T_0}^{T_1} dT_2 {T_2}^{5/2} E_{\rm H} \right] ^{-1/2} 
\end{split}
\label{eq8}
\end{align}

Using above equations, we derive the formulations of $E_{\rm H}$ and $T_{\rm max}$.
Assuming that $E_{\rm H}$ is uniformly distributed in a loop, and using $F_{\rm c}(T_{\rm max}) = 0$ in Equation (\ref{eq7}), we find
\begin{align}
E_H = \left( \frac{2}{2\alpha + 3} \right) \frac{7P^2}{8k_B^2} \chi T_{\rm max}^{\alpha -2}.
\label{eq9}
\end{align}
Substituting Equation (\ref{eq9}) into Equation (\ref{eq8}) yields
\begin{align}
l = \left[ \frac{\chi P^2}{(2\alpha +3)\kappa k_B^2} \right]^{-1/2} I_{2-\alpha} T_{\rm max}^{\frac{11-2\alpha}{4}},
\label{eq10}
\end{align}
and we find
\begin{align}
T_{\rm max} = \left[ \frac{\chi}{(2\alpha +3)\kappa k_B^2 I_{2-\alpha}^2} \right]^{\frac{2}{11-2\alpha}} (Pl)^{\frac{4}{11-2\alpha}}, 
\label{eq11}
\end{align}
where 
\begin{align}
I_{2-\alpha} \equiv \int_0^1 dt t^{\frac{7-2\alpha}{4}} \left(1-t^{2-\alpha}\right)^{-1/2}, \hspace {5mm}  t\equiv T/T_{\rm max}.
\label{eq12}
\end{align}
The relation for $E_H$ is derived using Equations (\ref{eq9}) and (\ref{eq11});
\begin{align}
E_H = \left( \frac{2}{2\alpha +3}\right)^{\frac{7}{11-2\alpha}} \left( \frac{\chi}{2\kappa k_B^2 I_{2-\alpha}^2}\right)^{\frac{2\alpha -4}{11-2\alpha}} \frac{7\chi}{8k_B^2} P^{\frac{14}{11-2\alpha}} l^{\frac{4\alpha -8}{11-2\alpha}}.
\label{eq13}
\end{align}
We note that $\alpha \neq 11/2$ is ensured from the form of $\Lambda (T)$; because we here treat the gas with at least $T>10^5$ K for deriving relations among the coronal quantities, Figure \ref{fig_fitz} illustrates $\alpha > 1$ (see also Table \ref{table_rad}).
Thus, given the values of $\alpha$ and $\chi$, the relations for loops with different elemental abundances can be derived in the forms of $T_{\rm max}=a(Pl)^b$ and $E_H = c P^d l^{e}$.
It can also be confirmed that $\alpha = -1/2$ and $\chi = 10^{-18.8}$ for the solar metallicity reproduce the original RTV scaling laws (Equations (\ref{eq1}) and (\ref{eq2})).

%\textcolor{red}{The derived scaling laws are used to understand the physical structure of a static coronal loop which can be regarded as a typical component of the averaged stellar atmosphere.
%In reality, however, coronal loops are known to show dynamic behaviors.
%We will discuss this issue in Section \ref{sec53}.}

\section{Simulation setup}
\label{sec3}

To study the dependence of metallicity on coronal properties, we perform MHD simulations for heating of a single coronal loop by explicitly incorporating dynamical processes, such as wave propagation, shock heating, and turbulence heating.
We verify the consistency with the derived scaling laws in Section \ref{sec2} by comparing the time-averaged coronal quantities from our simulations.

We also mention the updates from our previous numerical study on metalliciy dependence of coronal loops.
The differences from \cite{Washinoue2019} are the incorporation of the phenomenological turbulent heating, which is necessary to reproduce the thermally stable loop (Section \ref{sec1}), and the improvement of the modeling for chromospheric radiative loss functions which will be described in Section \ref{sec32}.

\subsection{Stellar model and loop models}
\label{sec31}

\begin{table}[t]
  \caption{Stellar parameters adopted in our simulations. Each row presents stellar mass, stellar radius, effective temperature, photospheric mass density and photospheric field strength.}
  \label{table_m08}
  \center
  \begin{tabular}{ccc} 
  \hline 
   & $M_{\star}$ [$M_{\odot}$]  & 0.8 \vspace{1mm}\\
   & $R_{\star}$ [$R_{\odot}$]  &   0.737 \vspace{1mm}\\
   & $T_{\rm eff}$ [K] & 5100 \vspace{1mm}\\
   & $\rho_{\rm ph}$ [g cm$^{-3}$] & 4.37 $\times 10^{-7}$ \vspace{1mm}\\
   & $B_{\rm ph}$ [kG]  &  1.96 \vspace{1mm}\\ 
   \hline
  \end{tabular}
\end{table}

We model a star with $M_{\star}=0.8M_{\odot}$, the age of $t=5$ Gyr and stellar metallicities of $Z=10^{0.5},10^{0}, 10^{-1}, 10^{-2}, 0 Z_{\odot}$, where $Z_{\odot}=0.014$ is the solar metallicity \citep{Asplund2009, Asplund2021}.
%$Z$ is basically varied by changing [Fe/H].
%The abundance of the other elements is determined by the solar abundance for $Z>Z_{\odot}$.
%For the lower-metallicity cases with $Z\leq 0.1 Z_{\odot}$, the primordial ratios are adopted \citep{Sutherland1993}. 
The basic parameters for $Z=Z_{\odot}$ are obtained from the stellar evolution calculation \citep{Yi2001, Yi2003} and the model of stellar atmospheres \citep{Kurucz1979} (Table \ref{table_m08}).

In this study, we consider a loop as a one-dimensional semicircular magnetic flux tube anchored in the photosphere, where we adopt the same configuration in \cite{Washinoue2022}.
The expansion factor $f(s)$ is given at each position to define a loop geometry.
$f_{\rm max}$, the value of $f(s)$ at the loop top, determines the coronal field strength $B_{\rm cor}$ by $B_{\rm cor}=B_{\rm ph}/f_{\rm max}$.
We adopt $f_{\rm max}=200$ to set $B_{\rm cor}=9.8$ G.
We use the same stellar and loop parameters for all the cases to easily compare the dependence of the coronal properties on metallicity.

\subsection{Equations}
\label{sec32}

We solve the one-dimensional MHD equations with radiative cooling, thermal conduction and phenomenological turbulent dissipation (see Section 2.3 in \cite{Washinoue2022}).
%\textcolor{red}{The differences from \cite{Washinoue2019} are the incorporation of the turbulent heating, which is necessary to reproduce the thermally stable loop (Section \ref{sec1}), and the improvement of the modeling for chromospheric radiative loss functions.}

We calculate the volumetric cooling rate, $q_{\rm R}$, by smoothly connecting radiation cooling in the chromosperic condition, $q_{\rm chrom}$, and in the coronal condition, $q_{\rm cor}$, in the following way: 
\begin{align}
q_{\rm R} = \xi q_{\rm chrom} + ( 1-\xi ) q_{\rm cor} \times \left( \frac{\rho}{\rho_{\rm bnd}}\right)^{-1} \mathrm{tanh}\left(\frac{\rho}{\rho_{\rm bnd}}\right), 
\label{eq14}
\end{align}
where
\begin{align}
 \xi = \mathrm{max} \left[0,\mathrm{min}\left[1,\frac{T_{\rm bnd}-T}{T_c}\right]\right].
\label{eq15}
\end{align}
In this study, we set $T_c = 5000$ K, $\rho_{\rm bnd}=10^{-13}$ g cm$^{-3}$ and $T_{\rm bnd}=3\times 10^4$ K.

We adopt the model for the chromospheric radiation from \cite{Carlsson2012}.
They developed a simple method to calculate the radiative loss and gain in the chromosphere using the results of the detailed radiative transfer calculation for the solar chromosphere.
In this model, the radiative cooling/heating rate is represented as the sum of the contribution from each element $X$;
\begin{align}
q_{\rm chrom} = \Sigma_X -L_{X_m}(T) E_{X_m}(\tau) \frac{N_{X_m}}{N_X}(T) A_X N_H n_e,
\label{eq16}
\end{align}
where $L_{X_m}(T)$ is the optically thin radiative loss function, $E_{X_m}(\tau)$ is the escaping probability, $\frac{N_{X_m}}{N_X}(T)$ is the fraction in ionization stage $m$, and $A_X$ is the abundance of element $X$.
In \cite{Carlsson2012}, the functions for H I, Ca II and Mg II are available.
The functional data for other stellar chromospheres with different abundances have not been provided yet. 
In this study, we change the metallic abundances by adjusting the values of $A_{\rm Ca}$ and $A_{\rm Mg}$ to apply Equation (\ref{eq16}) to the different metallicity cases.
The validity of this simplified treatment will be discussed in Section \ref{sec55}.

In the corona, the optically thin radiative cooling rate is simply calculated by 
\begin{align}
q_{\rm cor} = \Lambda(T) nn_e.
\label{eq17}
\end{align}
We utilize the optically thin cooling function $\Lambda(T)$ for different metallicities from \cite{Sutherland1993} (Figure \ref{fig_fitz}), where the functions are calculated assuming the collisional ionization equilibrium.
We note that since their original functions are based on the solar metallicity of $Z_{\odot}=0.019$ \citep{Anders1989}, we adjust the functions for the updated value of $Z_{\odot}=0.014$ \citep{Asplund2009, Asplund2021}.

\subsection{Initial conditions}

We set a uniform temperature distribution at $T=T_{\rm eff}$ as the initial condition.
We inject velocity fluctuations with the root-mean-squared amplitude of $\delta v = 1.0$ km s$^{-1}$ from the bottom boundaries (photospheres at $s=0$).
MHD waves (Alfv\'{e}n waves and acoustic waves) are generated by the footpoint motions.  
The frequency spectrum is proportional to $\omega_{-1}$ where we adopt $\omega$ from $\omega_{\rm min}=5.0 \times 10^{-4}$ Hz to $\omega_{\rm max}=3.3 \times 10^{-2}$ Hz.

\section{Simulation result}
\label{sec4}

\subsection{Time-averaged profiles}
\label{sec41}

\begin{figure}[t]
\begin{centering}
\includegraphics[width=9cm]{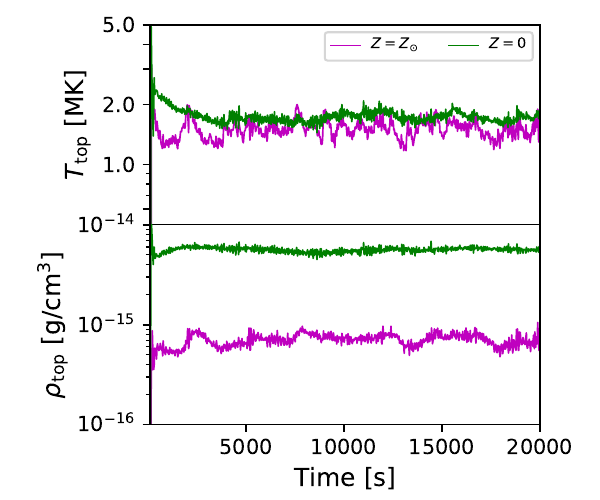}
\end{centering}
  \caption{Time evolution of the temperature (top) and density (bottom) at the loop top for $Z=Z_{\odot}$ (magenta) and $Z=0$ (green).
  }
  \label{fig_timevol}
\end{figure} 

\begin{figure}[!htb]
\begin{center}
\includegraphics{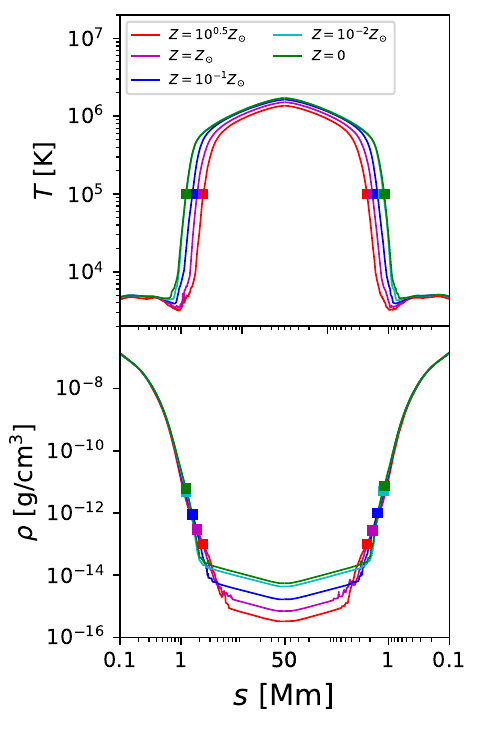}
\end{center}
  \caption{Time-averaged profiles of the temperature and density for different metallicities.
  The squares indicate the positions at the  transition region ($T=10^5$ K).}
  \label{fig_tzdep}
\end{figure} 

\begin{figure}[!t]
\begin{centering}
\includegraphics{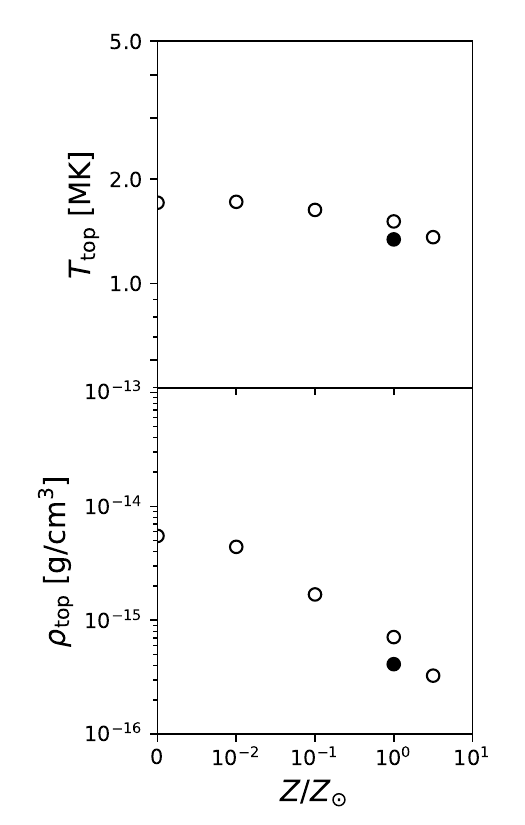}
\end{centering}
  \caption{Temperature (top) and density (bottom) at the loop top with metallicity for $l_{\rm loop}=50$ Mm. Filled circles are plotted for the case simulated with the coronal abundances.}
  \label{fig_top}
\end{figure} 

Once the simulations start, the isothermal loops are instantly heated to high temperatures to maintain the corona with $T>10^6$ K.
Figure \ref{fig_timevol} shows the time evolutions of the temperature and density at the loop top for $Z=Z_{\odot}$ and $Z=0$.
The major difference from the previous study is their steady-state behavior.
In \cite{Washinoue2019}, who do not include the turbulent dissipation, the loops repeat the cyclic evolution of the formation and destruction of high-temperature corona with time.
In addition to the intermittent shock heating, the turbulent dissipation provides continuous uniform heating and stabilize the coronal conditions, allowing us to clearly capture the dependence on metallicity.

The time-averaged loop profiles are shown in Figure \ref{fig_tzdep}.
We see that the coronal temperature is higher for lower metallicity because of the inefficiency of the radiative cooling.
In addition, the coronal density is considerably affected by metallicity.
This is because the gas with higher density can be heated up to a higher temperature owing to the smaller cooling efficiency in the lower-metallicity condition.
%This is because the gas with the higher density can be heated due to the smaller radiative cooling.}
Comparing the values of the density at the loop top, $\rho_{\rm top}$ (right panel of Figure \ref{fig_top}), lower-metal coronae with $Z \leq 0.01 Z_{\odot}$ give nearly one order magnitude larger $\rho_{\rm top}$ than metal-rich coronae with $Z>Z_{\odot}$, which is consistent with the results in the former works by \cite{Suzuki2018} and  \cite{Washinoue2019}.
The heating of the denser gas also lowers the location of the transition region at $T=10^5$ K.

\subsection{XUV radiations}
\label{sec42}

\begin{figure}[t]
\begin{centering}
\includegraphics{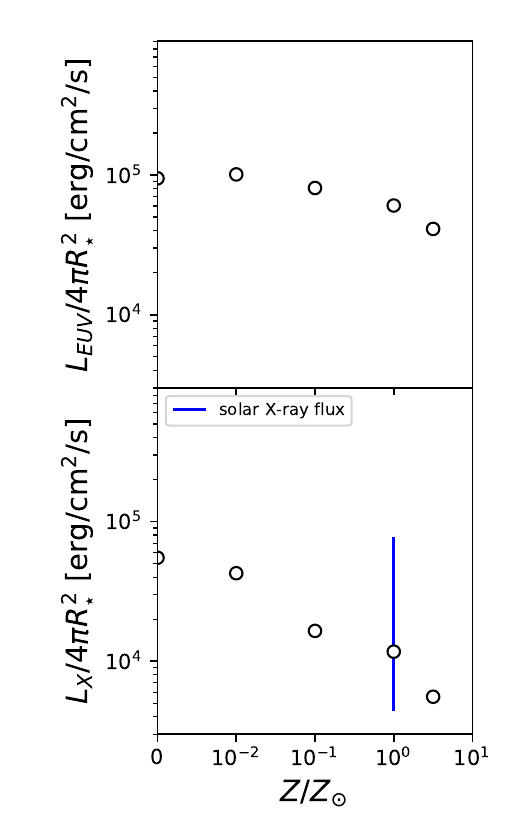}
\end{centering}
  \caption{EUV (left) and X-ray (right) fluxes with metallicity for $l_{\rm loop}=50$ Mm. The blue bar in the right panel shows the solar X-ray flux.
  }
  \label{fig_lxuv}
\end{figure} 

\begin{table}[!htb]
  \caption{Power-law index $\alpha$ and coefficient $\chi$ for the optically thin cooling function in $10^{5.7}$ [K] $\leq T\leq 10^{6.7}$ [K].}
  \label{table_rad}
  \centering
\begin{tabularx}{7cm}{@{\extracolsep{\fill}}YYYY}
  \hline
    Abundance & $Z/Z_{\odot}$ &  $\alpha$ & log$_{10}(\chi)$\\[1mm]
    \hline
    \hline
    & $10^{0.5}$  & -0.93  &  -16.03\\[1mm]
    & $10^0$  & -0.87 &  -16.85\\[1mm]
   photosphere & $10^{-1}$ & -0.71 &  -18.40\\[1mm]
    & $10^{-2}$ & -0.16 &  -22.21\\[1mm]
    & $0$  &  0.06 &  -23.64\\[1mm]
    \hline
    corona  & $10^0$  & -0.85 &  -16.39\\[1mm]
    \hline
  \end{tabularx}
\end{table}

\begin{table*}[t]
  \caption{The coefficients and indices in the scaling relations (Equations (\ref{eq19}) and (\ref{eq20})) calculated from Equations (\ref{eq11}) and (\ref{eq13}).}
\label{table_abcde}
  \centering
  \begin{tabularx}{14cm}{@{\extracolsep{\fill}}YYYYYYY}
  \hline
    Abundance & $Z/Z_{\odot}$ & $a$ & $b$ & $c$ & $d$ & $e$\\[1mm]
    \hline
    \hline
    & $10^{0.5}$  & 2670 &  0.311 & $7.00\times 10^5$ & 1.09 & -0.911 \\[1mm]
    & $10^0$  & 2080 &  0.314 & $3.08\times 10^5$ & 1.10 & -0.901\\[1mm]
    photosphere &$10^{-1}$ & 1360 &  0.322 & $7.47\times 10^4$ & 1.13 & -0.873\\[1mm]
    & $10^{-2}$ & 500 &  0.353 & $3.09\times 10^3$ & 1.24 & -0.763\\[1mm]
    & $0$  & 340 &  0.368 & $8.53\times 10^2$ & 1.29 & -0.713\\[2mm]
    \hline
    corona & $10^0$ & 2500 & 0.315 & $5.93\times 10^5$ & 1.10& -0.898 \\[1mm]
    \hline
  \end{tabularx}
\end{table*}

Subsequently, we estimate $L_{\rm EUV}$ and $L_{\rm X}$ from our results using the following equation;
\begin{align}
L = \frac{4\pi R_{\star}^2}{f_{\rm max}} \int q_{\rm R} f(s) ds,
\label{eq18}
\end{align}
where we assume that the stellar surface is filled with the simulated loops with a single size.
Figure \ref{fig_lxuv} shows the EUV and X-ray fluxes with different metallicities.
The blue bar in the right panel represents the solar X-ray flux.
%The value of $L_{\rm X}/4\pi R_{\star}^2$ for $Z=Z_{\odot}$ in the lower end of the bar is reasonable given the small surface area of the star with $M_{\star}=0.8 M_{\odot}$ in addition to the setup for the quiescent coronal heating.

It can be seen that $L_{\rm EUV}/4\pi R_{\star}^2$ shows a weak dependence on $Z$.
For the lower metallicity, the loops have a higher coronal density, which gives the larger emissivity. 
On the other hand, the spatial region that emits the EUV radiation is reduced because most of the gas in low-metal coronae is above the EUV temperature.
These counteracting effects result in a weak dependence of $L_{\rm EUV}$ on $Z$.

In contrast, the dependence of $L_{\rm X}/4\pi R_{\star}^2$ is found to be strong, owing to the larger spatial region with the X-ray temperature and the higher coronal density.
In particular, the X-ray flux for $Z=0$ is nearly one order magnitude larger than that for $Z=10^{0.5}Z_{\odot}$.
This trend explains the previous observational feature that some metal-poor stars emit strong X-rays despite their old age \citep[][]{Fleming1996, Ottmann1997}.

\section{Discussion}
\label{sec5}

\subsection{Comparison with the scaling laws}
\label{sec51}

\begin{figure*}
\begin{centering}
\includegraphics[width=170mm]{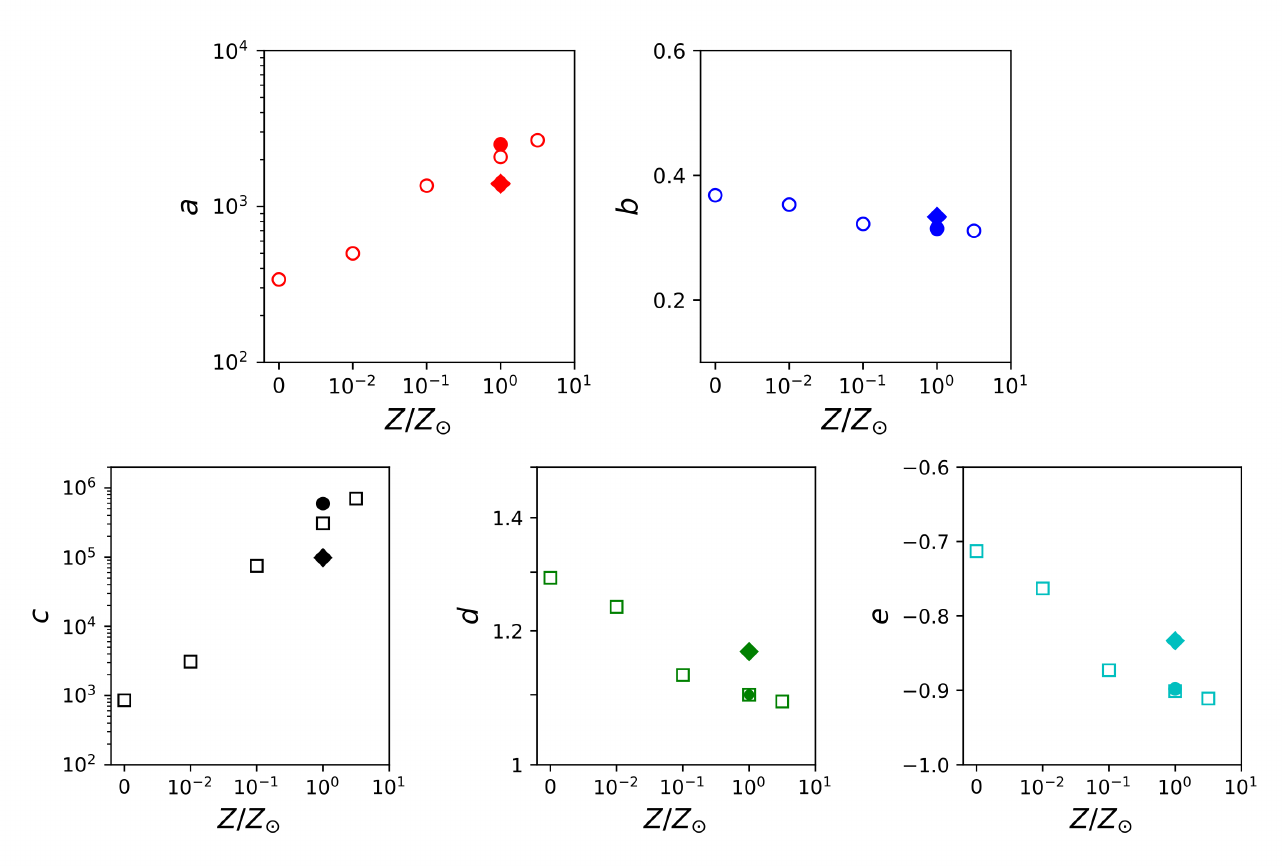}
\end{centering}
  \caption{The coefficients and indices of the scaling relations (Equations (\ref{eq11}) and (\ref{eq13})) with metallicity. The diamonds and filled circles are the values of the RTV scaling laws and for the coronal abundances, respectively.
  }
  \label{fig_abcde}
\end{figure*} 

\begin{figure*}
\begin{centering}
\includegraphics[width=170mm]{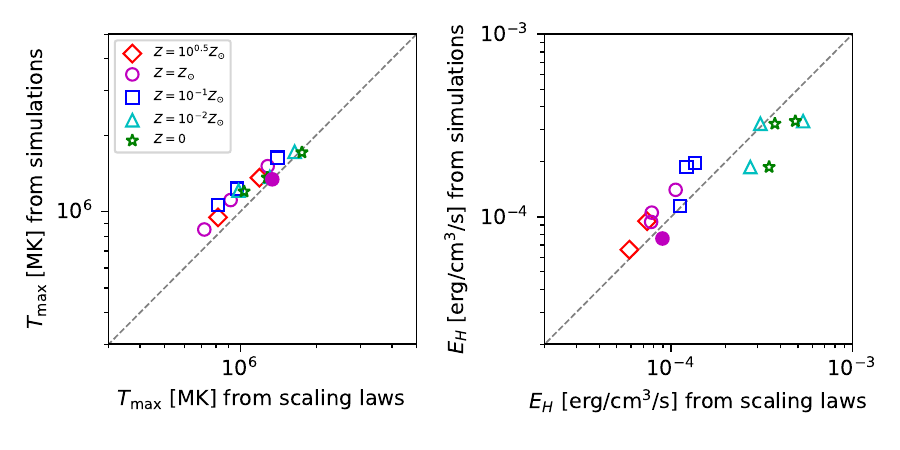}
\end{centering}
  \caption{$T_{\rm max}$ and $E_{\rm H}$ obtained from our simulations vs. those expected from the scaling laws (Equations (\ref{eq11}) and (\ref{eq13})). The filled symbols are the data from the simulation using the solar coronal abundances.
  }
  \label{fig_consistency}
\end{figure*} 

We compare our simulation results with the generalized scaling relations (Equations (\ref{eq11}) and (\ref{eq13})) derived in Section \ref{sec2}.
We adopt the single power-law fitting to the radiative loss functions (Equation (\ref{eq6})) for each metallicity over $10^{5.7}$ [K] $\leq T \leq 10^{6.7}$ [K] (dashed lines in Figure \ref{fig_fitz}).
In this work, we apply the narrow temperature range compared to  \cite{Rosner1978} where they consider the range of $10^{5.1}$ [K] $\leq T\leq 10^7$ [K]. 
The primary reason is to improve the accuracy of the power-law fitting.
In particular, a single power-law approximation over the broad temperature range is inappropriate for the cooling functions with $Z\leq 10^{-2}Z_{\odot}$ (Figure \ref{fig_fitz}).
Besides, since the gas with the temperature around a few $10^5$ K is thermally unstable ($\partial \Lambda / \partial T < 0$), we suspect that the assumption of the steady corona is hardly satisfied in the low temperature regime.

The values of $\alpha$ and $\chi$ obtained from our fitting are presented in Table \ref{table_rad}.
We mention that these depend on the form of $\Lambda(T)$. 
However, because the difference between the curves by \cite{Sutherland1993} and subsequent studies on radiative loss functions appears mainly in the low-temperature regime with $T \lesssim 10^5$ [K] \cite[e.g.,][]{Schure2009}, we expect that the values of $\alpha$ and $\chi$ are not significantly affected by the choice of $\Lambda(T)$ within the range of $10^{5.7}$[K]$<T<10^7$[K].
When we write the generalized scaling laws as 
\begin{align}
T_{\rm max} = a(Pl_{\rm col})^{b}
\label{eq19}
\end{align}
and 
\begin{align}
E_{\rm H} = c P^d {l_{\rm col}}^{e},
\label{eq20}
\end{align}
where $a,b,c,d$ and $e$ are calculated by using $\alpha$ and $\chi$ from Equations (\ref{eq11}) and (\ref{eq13}).
Table \ref{table_abcde} and Figure \ref{fig_abcde} show these values for different metallicities. 
In particular, a clear dependence on $Z$ is seen in the coefficients $a$ and $c$.
\footnote{Alternatively $a, b, c, d$ and $e$ can be directly fitted with $Z$. 
However, we do not employ this method because the multiple fitting with ($\alpha, \chi$) and ($a,b,c,d,e$) lowers the accuracy.}
We can interpret that, for the fixed $P$ and $l_{\rm cor}$, metal-poor corona gives smaller $T_{\rm max}$, because the heating required to balance the smaller radiative loss is also smaller (Equation (\ref{eq19})).
On the other hand, when the same energy is input as in our simulations, the coronal pressure is considerably higher for the lower-metallicity loop because of the suppression of the radiative cooling (Equation (\ref{eq20})).
Therefore, in our simulations, metal-poor coronae show higher maximum temperatures.
%Moreover, the elemental composition changes with height from the photosphere to the corona as we will discuss in Section \ref{sec52}.
%Thus, strictly speaking, $\Lambda(T)$ should not be characterized with $Z$ alone.}}

The filled circles in Figure represent the values for the RTV scaling laws (Equations (\ref{eq1}) and (\ref{eq2})).
The scaling laws for the solar coronal loop is slightly modified due to the different radiative loss function and the narrower fitted range of the temperature.
Meanwhile, $T_{\rm max}$ and $E_{\rm H}$ for fixed $P$ and $l_{\rm cor}$ do not yield large differences from those calculated by the original RTV scaling laws; when $P=1$ dyne cm$^{-2}$ and $l_{\rm cor}=10^{9}$ cm, our derived relations give $T_{\rm max} = 1.39 \times 10^6$ K and $E_{\rm H}= 2.40 \times 10^{-3}$ erg cm$^{-3}$ s$^{-1}$, while the original RTV scaling laws give $T_{\rm max} = 1.40 \times 10^6$ K and $E_{\rm H}= 2.97 \times 10^{-3}$ erg cm$^{-3}$ s$^{-1}$.

Figure \ref{fig_consistency} compares the simulation results and the derived scaling laws, where on the vertical axis $T_{\rm max}$ (left) and $E_{\rm H}$ (right) are directly measured from the simulations and on the horizontal axis they are calculated from Equations (\ref{eq11}) and (\ref{eq13}) using $P$ and $l_{\rm cor}$.
It can be seen that our simulations are in good agreement with the scaling relations for each metallicity.
The assumption of the constant pressure over a loop, which is adopted in deriving the scaling laws, is not in fact satisfied in MHD simulations. This is because the simulated loop has a pressure gradient, resulting in a loop-averaged pressure smaller than that used in the scaling laws. 
Nevertheless, because the total pressure is increased by turbulent and magnetic pressure, the simulations show a close agreement with the scaling laws. 
As our MHD simulations can treat the dynamical processes which work in realistic stellar atmosphere, these results support the applicability of the scaling relations to observed stars.

\subsection{Elemental abundances in the corona}
\label{sec52}

\begin{figure}[t]
\begin{centering}
\includegraphics[width=8.5cm]{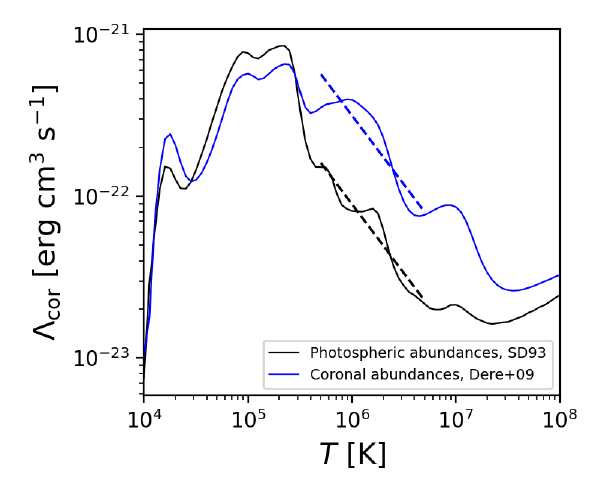}
\end{centering}
  \caption{Optically thin radiative loss functions for the solar photospheric (black) and coronal (blue) abundances. The dashed lines are the fitting lines over $10^{5.7}$ [K] $\leq T \leq 10^{6.7}$ [K].
  }
  \label{fig_corcol}
\end{figure} 

\begin{figure}[t]
\begin{centering}
\includegraphics{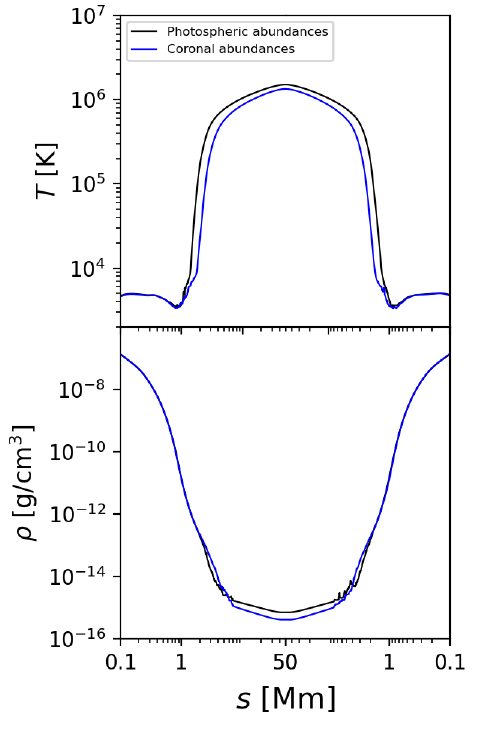}
\end{centering}
  \caption{Time-averaged profiles of the temperature and density for the photospheric (black) and coronal (blue) abundances.
  }
  \label{fig_compthin}
\end{figure} 

%\textcolor{red}{In this section, we investigate the effect of abundance anomalies observed in the stellar atmosphere on the scaling relations and  coronal properties by performing numerical simulations for the solar case.}
In our simulation, we adopt the cooling functions for the photospheric abundances.
In fact, however, the elemental abundances in the corona are known to be different from those in the photosphere.
In the solar corona, the elements with a low first ionization potential $< 10$ eV are more abundant than in the photosphere, which is called the FIP effect \citep{Pottasch1963}.
Similar abundance anomaly has been detected in the stellar coronae with the low-activity level \citep{Drake1997}.
In this section, we examine how the coronal properties change when the coronal abundances are employed, compared to when we adopt the photospheric abundances.

For the Sun, the radiative loss function with the coronal abundances is calculated by \cite{Dere2009}, where the coronal abundances are adopted from \cite{Feldman1992}.
The cooling functions with the photospheric and coronal abundances are compared in Figure \ref{fig_corcol}.
The large deviation is seen in $10^{5.7}$ [K] $\leq T \leq 10^7$ [K] where the low-FIP elements such as Fe and Si make large contributions to the total cooling.
Therefore, it is expected that the coronal properties are affected by which elemental abundances are assumed.

We fit the cooling function for the solar coronal abundances to the single power law as in Section \ref{sec41} (the blue dashed line in Figure \ref{fig_corcol}).
The indices and coefficients in the fitting function and the scaling relations are presented in Tables \ref{table_rad}, \ref{table_abcde}, and Figure \ref{fig_abcde}.
For the coronal abundances, the power-law index of the cooling function $\alpha$ is almost the same as the function for the photospheric abundances.
Therefore, only the coefficients in the scaling relations, $a$ and $c$, are substantially changed.
For the fixed $P$ and $l_{\rm cor},$ $T_{\rm max}$ and $E_{\rm H}$ are increased by a factor of $1.2$ and $1.9$ compared to the case for the photospheric abundances.

In order to compare the atmospheric profiles between the cases of different abundances, we additionally carried out the simulations to compare the coronal atmospheres with the solar photospheric and coronal abundances.
Figure \ref{fig_compthin} shows the time-averaged profiles of the temperature and the density.
It is found that when the coronal abundances are employed, the coronal temperature and density are reduced due to the enhancement of the metallic cooling.
We also confirm that $T_{\rm max}$ and $E_{\rm H}$ obtained from our simulations are consistent with those calculated from the generalized scaling relations.
Our results indicate that we should be careful about the elemental abundances in the atmosphere to specifically reproduce stellar coronal heating by simulations.
%$T_{\rm top}$ and $\rho_{\rm top}$ are plotted in Figure \ref{fig_ztop} with the filled circles. 

We also note the inverse FIP (IFIP) effect. 
In the corona of some magnetically active stars low-FIP elements are less abundant than in the photosphere, which shows the depleted low-FIP elements in the corona compared to the photosphere \citep{Brinkman2001, Robrade2005}.
Recently, the IFIP effect is also detected in solar flares \citep{Doschek2016, Katsuda2020} and slow solar winds \citep{Brooks2022}.
The observations have revealed that the elemental abundances in the stellar coronae depend on their activity; the low-FIP elements are rich in inactive stars, and deficient in active stars \citep{Drake2002}.
These observations indicate that the abundance ratios of low-FIP elements to high-FIP elements in the corona depend on different regions even in a single star.
Therefore, we need to caution the variety of (I)FIP effect in studying the coronal heating in different-metallicity stars. 

\subsection{Application of the scaling relations to observations}
\label{sec53}

Our derived scaling laws can be used as alternative relations to the RTV scaling laws for the solar corona to compare with observations of stars with different metallicities, while it is difficult to apply to loops on other stars because they are generally observed as point sources.
However, even for the solar corona, it is known that some loops observed in active regions often do not satisfy the RTV scaling laws \citep{Aschwanden2008}.
This is because most of active-region loops are thermally unstable, and do not satisfy the assumption of a static state in deriving RTV scaling laws.
Nevertheless, it is still useful to diagnose the dynamical nature of observed loops and estimate the heating distribution by measuring deviations from the theoretical prediction.
In addition, \cite{Aschwanden2008b} applied the RTV scaling laws to construct the relations for estimation of the electron density in flaring loops, which are confirmed to be consistent with observations.
For the solar corona, the physical relations for dynamic loops have been derived as the extended RTV scaling laws \citep{Bradshaw2020} and the zero-dimensional model of loop evolution \citep{Cargill2012}.
It will be also useful to extend our scaling laws to those for dynamic loops to study various atmospheric structures.

For stellar coronae, the scaling laws can also be applied to the analysis of very long loops which have sometimes detected in active stars \citep[e.g.][]{Peterson2010}.
In addition, the derived scaling laws will also be utilized to estimate the physical quantities that cannot be resolved by observations.
A similar attempt has been done for solar and stellar flares by \cite{Namekata2017}; they estimate the coronal field strength and loop length using the observed correlation between the emission measure and temperature of flares \citep{Shibata2002}.
Using Equations (\ref{eq11}) and (\ref{eq13}) and the observable quantities of the coronal temperature, density and metallicity, we will be able to estimate the typical loop length.
At this moment, the number of the detection of metal-poor coronae is still limited probably because metal-poor stars are magnetically inactive owing to their age.
\cite{Ottmann1997} surveyed the X-ray emission from Population II binaries to clarify the effect of metallicity on the coronal activities by ROSAT observations. 
They reported that the average X-ray luminosities from the metal-poor coronae are low compared to those from the Population I binaries. 
They mention the possibility that the less activity in the metal-poor
coronae is associated with the loss of the convective envelope due to their old age.
In contrast to these results, strong X-ray emission has also been reported in a few old metal-poor stars \citep{Fleming1996, Ottmann1997, Guinan2016}, which can be explained by our simulation results.
Thus, to understand statistical properties of stellar coronae, further extensive studies are required..
Large surveys with future observations will reveal the atmospheric structures with various metallicities.
In particular, the X-Ray imaging spectroscopy mission (XRISM), which is expected to be launched in 2023, will provide large samples of metal-poor stars with X-ray coronae with high-resolution spectroscopic measurement.
The generalized scaling relations will be then useful to estimate the coronal quantities and assess the dynamical nature of the atmosphere.

\subsection{Dependence of input energy flux on stellar metallicity }
\label{sec54}

In this study, we only change the stellar metallicity without taking into account the dependence of the photospheric parameters on metallicity (Section \ref{sec31}).
It enables us to clearly understand the importance of the cooling efficiency in coronal heating, however, the photospheric condition actually depends on metallicity.
In particular, the magnitude of velocity perturbation $\delta v$ is a critical parameter which controls the input Poynting flux from the photosphere.
Although it is hard to uniquely determine the value of $\delta v$ because of no constraints from observations, we possibly estimate it with a theoretical approach.

\cite{Musielak2002} analytically investigated the effect of metallicity on the energy flux of transverse waves generated from the surface convection.
They show that for the stars with $T_{\rm eff} < 6000$ K, the energy flux decreases with decreasing metallicity.
According to their computation, the energy flux is one order magnitude smaller when metallicity is decreased by one order magnitude for stars with the same $T_{\rm eff}$.
%Since the opacity is smaller for lower-metal stars, the photosphere tends to be located at a deeper position.
From the relation of $\rho \delta v^3 \approx \sigma T_{\rm eff}^4$ \citep[e.g.,][]{Bohn1984}, the value of $\delta v$ for the stars with the same mass is smaller for lower metallicity even when the larger $T_{\rm eff}$ for lower metallicity is taken into account.
Therefore, our simulations probably overestimate the input energy flux for lower-metallicity stars.
It needs to further investigate the dependence of coronal heating on metallicity using different values of $\delta v$ for each metallicity.

\subsection{Modeling the chromosphere with different metallicities}
\label{sec55}

While the model of the solar chromosphere has been improving, the applications to other stellar chromospheres focusing on the difference in the chemical composition has not been progressed.
The reliable descriptions for metal-rich and metal-poor chromospheres need to perform the direct radiation MHD simulations.
In this study, we only tune the metallic abundances ($A_X$) to include the effect of metallicity for the calculation of the radiative loss (Section \ref{sec32}).
However, in Equation (\ref{eq16}), the escaping probability $E_{X_m}(\tau)$ and the ionization fraction $\frac{N_{X_m}}{N_X}(T)$ are also expected to depend on metallicity.

The recipe from \cite{Carlsson2012} includes the above functional data for H I, Ca II and Mg II.
Since $A_{\rm Ca}$ and $A_{\rm Mg}$ are assumed to be linearly dependent on $Z$, the contribution of the metals to the chromospheric radiation is substantially reduced for $Z\le 0.1Z_{\odot}$. 
Therefore, the uncertainty of the functions for Ca II and Mg II under the metal-deficient environment is not a critical matter, although only that for hydrogen still remains as a concern.
For the higher metallicity with $Z > Z_{\odot}$, the cooling from the metallic lines is more effective.
Furthermore, the recipe adopted in our simulations does not incorporate the contribution from the iron lines, which are likely to be important in the chromospheric radiation \citep{Anderson1989}.
Therefore, Fe II should be also added to calculate the radiative loss, particularly for high-metallicity cases.
The validity of the present treatment to include the different metallicity needs to be examined with radiation transfer calculations.

Another missing physics in our chromospheric modeling is the magnetic diffusion caused by the ion-neutral interactions. 
Some numerical studies for the solar chromosphere demonstrate that the ambipolar diffusion plays an important role in the local heating in the chromosphere \citep{Khomenko2012, Martinez-Sykora2017,Nobrega-Siverio2020}, although the time-averaged heating rate by the ambipolar diffusion is found to be much smaller than shock heating rate from the nonlinear mode conversion of the Alfv\'{e}n waves \citep{Arber2016}.
Because of the lower ion density, the ambipolar diffusion is expected to be more effective in metal-poor chromospheres, which may affects the chromospheric structure and the coronal heating.
To clarify how the metallicity changes the chromospheric profiles, it is needed to incorporate the ambipolar diffusion term into the basic equations in our future work.

\begin{figure}[t]
\begin{centering}
\includegraphics[width=8.5cm]{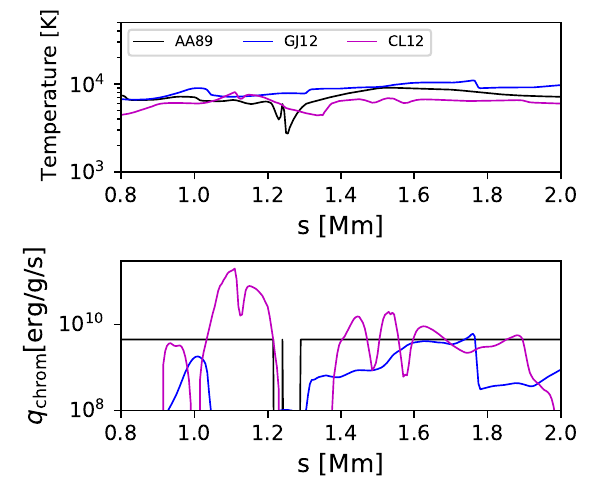}
\end{centering}
  \caption{Snapshots of the temperature structures (top) and the radiative cooling rate per mass (bottom) in the chromosphere for the different models.
  }
  \label{fig_comp}
\end{figure} 

\begin{figure}[t]
\begin{centering}
\includegraphics[width=8.5cm]{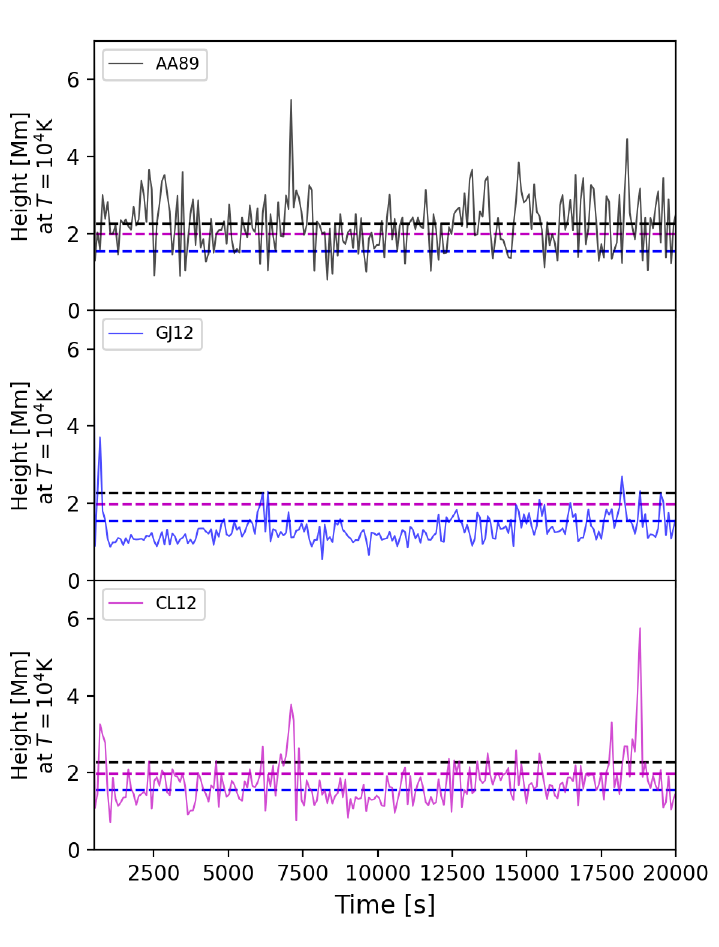}
\end{centering}
  \caption{Time evolution of the height at $T=10^4$ K for AA89 (top and black), GJ12 (middle and blue) and CL12 (bottom and magenta) models.The three dashed lines in each panel represent the time-average heights of these three models.
  }
  \label{fig_timech}
\end{figure} 

\begin{figure}[!htb]
\begin{centering}
\includegraphics[width=8.5cm]{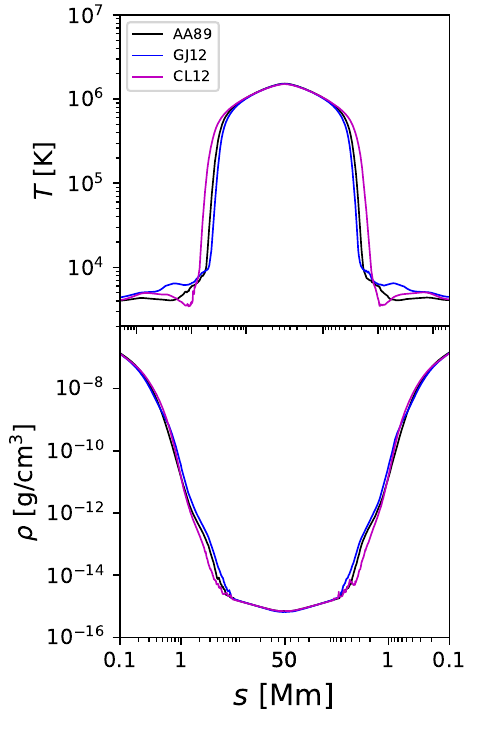}
\end{centering}
  \caption{Time-averaged profiles of the temperature and density for different chromospheric models.
  }
  \label{fig_aagjcl50}
\end{figure}

\section{Summary}
\label{sec6}

To construct the model of coronal loops with various metallicities, we generalized the scaling relations for a hydrostatic coronal loop from the energy balance among heating, radiative cooling and thermal conduction, as an extension of the RTV scaling laws for the Sun.
These allow us to characterize the coronal-loop structure with a variety of abundances once the optically thin radiative loss function for the plasma gas is obtained.
Our development of scaling relations enables the interpretation of data that will be obtained from future observations (e.g., XRISM). 
We expect that the derived scaling relations will be utilized to diagnose the physical quantities of coronal atmospheres for stars across a wide range of metallicity. 

The effect of the metallicity on coronal heating is investigated by the MHD simulations with radiative loss functions for different metallicities.
It is revealed that the coronal temperature and density are higher for lower metallicity because the radiative cooling is suppressed.
Accordingly, our estimates for the XUV luminosities give the larger values for lower-metallicity coronae.
Notably, the X-ray luminosity shows a steep dependence on metallicity.
This is because the coronal density sensitively depends on the metallicity; even though the cooling efficiency is lower for lower-metallicity cases, a significant increase in the coronal density surpasses its effect, yielding a larger value of $L_{\rm X}$.
These results on the coronal radiation would have a large impact on the evolution of protoplanetary disks and planetary atmospheres through gas evaporation.

Moreover, we discussed the (I)FIP effect in coronal plasmas.
A number of observations have reported that the elemental abundances in the corona are different from those in the photosphere, which depends on the stellar activity.
We simulated the coronal heating with the radiative loss functions for the solar photospheric and coronal abundances.
When we employ the coronal abundances, the increased coolants of low-FIP elements enhance the radiative cooling, which yields the cooler coronal atmosphere.
This indicates that conversely when the low-FIP elements are depleted as observed in magnetically active stars, the coronal temperature and density are expected to increase by a factor of a few.
Therefore, we suggest that a variety of elemental abundances should be taken into account for stellar coronal modelings.

Although the metallicity is found to play an essential role in controlling the coronal properties, there still remains challenges on modeling the stellar chromosphere with different metallicities.
Because we adopted the models based on those for the solar chromospheric conditions in calculating the radiative loss, the detailed processes of radiative transfer need to be investigated under different-metallicity conditions.
In addition, magnetic diffusion in partially ionized gas is an important process for the chromospheric heating.
To develop full pictures of the stellar atmosphere, further studies will be needed to assess these physical processes and the dependence of their effects on metallicity.

Numerical computations were in part carried out on PC cluster at Center for Computational Astrophysics, National Astronomical Observatory of Japan. 
H.W. is supported by JSPS KAKENHI grant No. JP22J13525.
T.K.S. is supported in part by Grants-in-Aid for Scientific Research from the MEXT/JSPS of Japan, 17H01105, 21H00033 and 22H01263 and by Program for Promoting Research on the Supercomputer Fugaku by the RIKEN Center for Computational Science (Toward a unified view of the universe: from large-scale structures to planets; grant 20351188-PI J. Makino) from the MEXT of Japan.

\appendix
\section{Radiative loss in the solar chromosphere}

One should be careful to model the radiative loss in the chromosphere, because the physical condition of the chromosphere is directly connected to the Alfv\'{e}n-wave propagation and the coronal heating \citep{Washinoue2022}.
Therefore, it is worthwhile to examine how the atmospheric profiles are affected by different models for the chromosphere.
Our simulations in this paper use the prescription by \cite{Carlsson2012} and extend it to different metallic abundances.
In this subsection, we compare the atmospheres with the following three models for the chromospheric radiative loss.

\noindent (I) AA89 model 

\cite{Anderson1989} introduced a chromospheric model with non-LTE effects to reproduce the temperature structure of the VAL C atmosphere \citep{Vernazza1981}.
They found that the volumetric cooling rate is almost proportional to the mass density in the temperature plateau with $T\lesssim10^4$ K which accounts for most of the VAL chromosphere:
\begin{align}
q_{\rm chrom} = 4.5\times 10^9 \rho .
\label{eqa1}
\end{align}
Owing to the simple form, Equation (\ref{eqa1}) has been widely used in numerical simulations that treat from the photosphere and the chromosphere to the corona \citep{Moriyasu2004,Suzuki2005,Matsumoto2010a,Suzuki2018,Washinoue2019}.
However, it should be noted that it is required to set a cutoff temperature of radiative cooling to avoid the extremely low chromospheric temperatures \citep{Washinoue2022}.

\vskip\baselineskip
\noindent (II) GJ12 model 

More precise modeling for the chromospheric radiative loss is developed by \cite{Goodman2012}.
They use the optically thin approximation, following that the metallic lines of Fe II, Ca II and Mg II dominate over the chromospheric radiation.
The radiative cooling rate is fitted to a three-level generic hydrogen atom with two excited states using the CHIANTI atomic database and the OPACITY project;
\begin{align}
q_{\rm chrom} = 8.63\times 10^{-6} C_E \frac{n_e (n_H + n_p)}{T^{1/2}} \sum_{i=1}^2 E_i \Gamma_i \mathrm{exp}\left( \frac{-eE_i}{k_B T}\right),
\label{eqa2}
\end{align}
where $C_E = 1.6022\times 10^{-12}$ erg eV$^{-1}$, the energies of a hydrogen atom with the excited state $E_1 = 3.54$ eV and $E_2 = 8.28$ eV.
$\Gamma_1 = 0.15\times 10^{-3}$ and $\Gamma_2 = 0.065$.
Equation (\ref{eqa2}) is confirmed to be valid in $T< 1.5\times 10^4$ K.
$n_e$ is calculated by the analytic expression derived for non-LTE hydrogen plasma in the solar photosphere and chromosphere:
\begin{align}
n_e = n_H^{1/2} \left( \frac{3.2\times 10^4}{\alpha}\right)^{1/2} \left( \frac{T}{10^4}\right)^{0.393} \mathrm{exp}(-T_2/2T),
\label{eqa3}
\end{align}
where $\alpha = 3.373\times 10^{-13}$ and $T_2 \sim 1.18187\times 10^5$ K.

\vskip\baselineskip
\noindent (III) CL12 model 

The model by \cite{Carlsson2012} is adopted in our simulations.
The cooling/heating rate is described as
\begin{align}
q_{\rm chrom} = \sum_X -L_{X_m}(T)E_{X_m}(\tau)\frac{N_{X_m}}{N_X}(T)A_X N_H n_e,
\end{align}
where each factor is already explained in Section \ref{sec32}.
This recipe allows us to compute the radiative loss and gain in the solar chromosphere with a low numerical cost, which well reproduces those obtained from the direct radiative transfer calculations.
This treatment can be used in the range of $T\leq 3\times 10^4$ K.
%where the coronal approximation is not justified.

\vskip\baselineskip
Figure \ref{fig_comp} compares snapshots of the temperature and the radiative cooling rate per mass for the three models, where we pick out a region that corresponds to the temperature plateau above the temperature-minimum region.
The AA89 model shows the constant cooling rate per mass above the cutoff temperature $T_{\rm off}=5000$ K.
On the other hand, for the GJ12 and CL12 models, $q_{\rm chrom}$ takes different values at different locations since it is calculated by the local density and temperature.
The time evolution of the height at $T=10^4$ K is shown in Figure \ref{fig_timech}.
In all the cases, this height dynamically moves up and down with time.
The dashed lines represent the time-averaged heights.
The difference among the three cases is less than 0.5 Mm.

The time-averaged structures of the temperature and density are compared in Figure \ref{fig_aagjcl50}.
There are some differences below the coronal base ($T=5\times 10^5$ K), while the coronal temperature and density are almost the same.
The AA89 and GJ12 models show a gradual increase in the chromospheric temperature.
On the other hand, since the significant local cooling frequently occurs in the CL12 models (Figure \ref{fig_comp}), the thinner chromospheres are formed due to the low $T_{\rm min}$.
%Among these models, the CL12 model, which is empirically obtained by direct radiation MHD simulations, is appropriate to accurately handle the dynamic behavior of the chromosphere.
%Therefore, when we employ the simplified model for the chromospheric radiative loss, we should note that there appears a slight differences in the chromospheric structure with different models.

\nocite{*}
\bibliography{washi23}

\end{document}